\documentclass[sigplan,screen, nonacm=true, authorversion=true]{acmart}

\AtBeginDocument{%
  \providecommand\BibTeX{{%
    \normalfont B\kern-0.5em{\scshape i\kern-0.25em b}\kern-0.8em\TeX}}}

\setcopyright{none}
\renewcommand\footnotetextcopyrightpermission[1]{}

\usepackage{graphicx}
\usepackage{subfigure}
\usepackage{xcolor}
\usepackage{xspace}
\usepackage{url}
\usepackage{breakurl}

\usepackage[utf8]{inputenc}
\usepackage[T1]{fontenc}
\usepackage{lmodern}
\usepackage[russian, english]{babel}

\begin{document}

\title{\vspace{-1em}\archshort: The case for a \arch for Serverless Computing}


\author{Anjo Vahldiek-Oberwagner}
\affiliation{%
  \institution{Intel Labs}
  \city{Berlin}
  \country{Germany}}

\author{Mona Vij}
\affiliation{%
  \institution{Intel Labs}
  \city{Hillsboro}
  \country{USA}}

\renewcommand{\shortauthors}{Vahldiek-Oberwagner and Vij}

\newcommand{\arch}{Memory-Safe Software and Hardware Architecture~}

\newcommand{\archshort}{MeSHwA\xspace}
\newcommand{\intel}{Intel\textsuperscript{\textregistered}\xspace}
\newcommand{\intels}{Intel's\textsuperscript{\textregistered}\xspace}

\newcommand{\todo}[1]{{\color{red}#1}}

\begin{abstract}
  Motivated by developer productivity, serverless computing, and microservices
have become the de facto development model in the cloud. Microservices decompose
monolithic applications into separate functional units deployed individually.
This deployment model, however, costs CSPs a large infrastructure tax of more
than 25\%~\cite{warehousescalecomputing, fbcommunication}. To overcome these
limitations, CSPs shift workloads to Infrastructure Processing Units (IPUs) like
Amazon's Nitro~\cite{amazon-nitro} or, complementary, innovate by building on
memory-safe languages and novel software
abstractions~\cite{fastly-wasm,cloudflare-workers}.

Based on these trends, we hypothesize a \arch providing a general-purpose
runtime environment to specialize functionality when needed and strongly isolate
components. To achieve this goal, we investigate building a single address space
OS or a multi-application library OS, possible hardware implications, and
demonstrate their capabilities, drawbacks and requirements. The goal is to bring
the advantages to all application workloads including legacy and memory-unsafe
applications, and analyze how hardware may improve the efficiency and security.

\end{abstract}

\maketitle

\section{Introduction}

Serverless computing, and microservices have recently gained in popularity to
deploy cloud services. They improve developer productivity by focusing on
business logic and separating functional services. Function-as-a-Service (FaaS)
is a special form of serverless computing. In an extreme case microservices
implement functional decomposition similar to FaaS. With FaaS offering the
smallest services, the techniques are on a spectrum for service size. Throughout this
paper we will refer to serverless computing, microservices and FaaS
interchangeably and refer to the application as a service. Cloud Service
Providers (CSPs) run and provide the infrastructure software connecting each
service. This abstraction comes at the cost of about 25\% infrastructure tax as
reported by Google~\cite{warehousescalecomputing, fbcommunication}, consisting
of various overheads due to small units communicating frequently over network.
Large CSPs react by shifting the infrastructure software into Infrastructure
Processing Units (IPUs), which operate more cost effectively than traditional
server CPUs offering higher performance for workloads. While this shifts the
burden and reduces cost, this approach does not reduce the communication cost,
or provide a software architecture eliminating the inherent overheads associated
with this deployment model. Alternatively, CSPs built special purpose
environments relying on memory-safe
languages~\cite{fastly-wasm,cloudflare-workers} which require specific compiler
and runtime environments and deny their optimizations to legacy applications.

To avoid the inherent overheads in today's hardware and software architecture,
we suggest leveraging memory-safe languages for performance and investigate
hardware optimizations generalize the environment. Memory-safe languages (e.g.,
Rust) and runtimes (e.g., Wasm) rely on the compiler to generate binary code
preventing memory access violations via static and runtime checks. Their
performance is comparable to native execution in case of Rust~\cite{perfrust}
and shows moderate overheads for Webassembly (Wasm)~\cite{perfwasm}. Serverless
computing could benefit from this environment by running all services in the
same process with near-zero communication cost while being isolated from each
other. At the same time, the runtime specializes functionality to, e.g., access
devices avoiding OS overheads.

These two infliction points, the advance in serverless computing and memory-safe
languages, suggests that we should revisit and explore a memory-safe software
and hardware architecture providing a general-purpose runtime environment to
specialize functionality when needed and strongly isolate services. Our goal is
to leverage memory-safe software guarantees where possible, and describe the
required hardware to further improve performance and security. The solution
needs to generalize to legacy and memory-safe services and offer easy adoption.
We will analyze the path of a new Single Address Space Operating System (SASOS)
using Rust and Wasm, as well as a library Operating System (library OS) allowing
multiple services to execute in parallel while also depending on a traditional
host OS. While this idea is not new and was tried in industry~\cite{midori,
singularity}, we believe the serverless and other cloud workloads demand a
drastic change in hardware and software.

\begin{figure*}[t]
    \includegraphics*[]{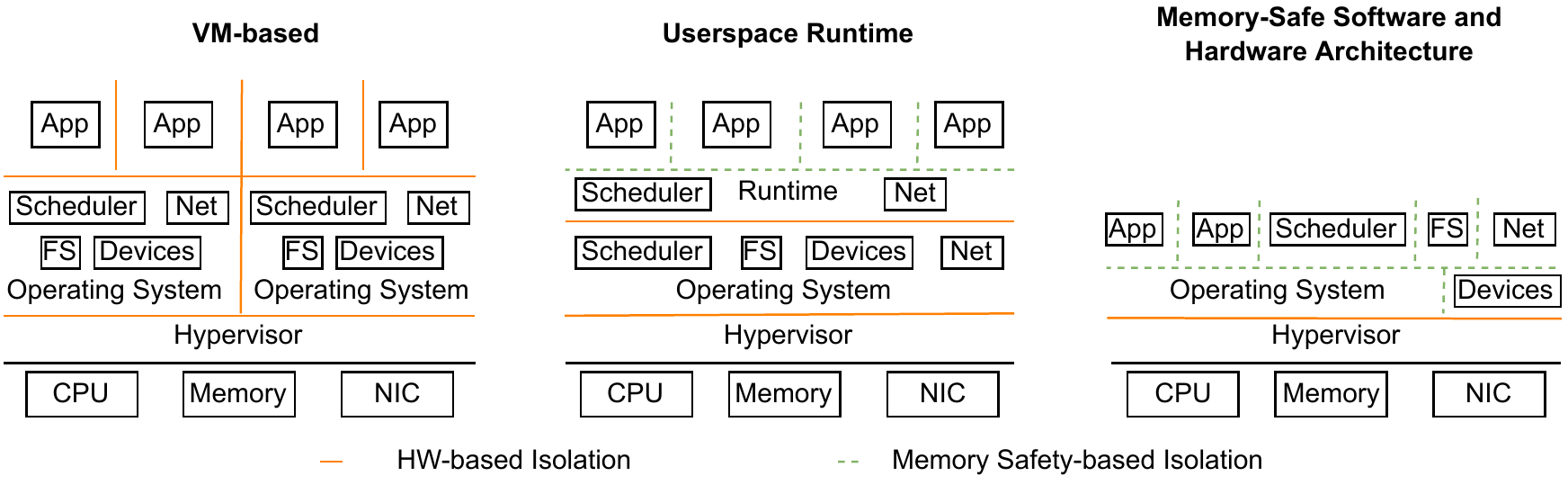}
    \caption{Comparison between cloud software architecture}
    \label{fig:mssa_comparison}
\end{figure*}

By leveraging memory-safe languages our architecture can benefit from a single
memory abstraction to drastically improve switching and communication between
services. This is in harsh contrast to traditional hardware and software
abstractions which rigidly isolate processes, operating systems (OS), address
spaces, and privilege levels. To achieve this goal, services and OS
functionality execute in the same address spaces and privilege level. Sharing
becomes instant between a service and the OS. Existing research~\cite{lwc, smv,
secage, nestedkernel, dune} on improving process-based isolation suffers from
too high runtime overheads. Research has focused on SASOS~\cite{singularity} and
library OSes ~\cite{demikernel, graphene, ix, arrakis, redleaf,akkus2018sand} as
the two alternative approaches to bridge this gap. SASOS overcome existing
inefficiencies in traditional OSes by executing the OS and the service in the
same address space while offering modularity. Recent advances in memory-safe
languages and runtimes offer the building blocks to restart SASOS efforts, but
require extensive engineering effort to build a new OS and further improvements
to secure memory-safe environments to prevent advanced attacks (e.g., transient
execution attacks~\cite{narayan2021swivel}). In contrast, library OSes move
kernel functionality into the userspace and offer specialized alternatives like
kernel-bypasses for networking~\cite{demikernel}. Instead of writing an OS from
scratch, library OSes focus on a specific technique and use the host OS to
support the remaining functionality. These approaches typically lack the support
for running multiple services at the same time while properly isolating their
execution.

We discuss the main beneficiaries of \archshort including microservices, FaaS
and memory-bandwidth intensive workloads such as machine learning. An
inefficient software architecture causes the infrastructure tax to be up to 25\%
~\cite{warehousescalecomputing}. The remainder of this paper discusses how to
build and optimize \archshort for such workloads.
\section{Background}

\paragraph{Modern Cloud Deployments}
Server applications increasingly replicate functionality traditionally found in
operating systems. Their intent is to improve performance by specializing
functionality to the workload, and provide a portable environment across
operating systems and hardware. In case of Fastly and
Cloudflare~\cite{fastly-wasm,cloudflare-workers} most functionality is executed
in userspace reducing context switch overheads and infrastructure tax. These
providers achieve this goal by executing functions in the services together with
a userspace scheduler, memory management, and dedicated userspace network
stacks. Library OS~\cite{demikernel, graphene, ix, arrakis} similarly improve
performance of applications via reducing system calls and bypassing the kernel
for network and storage requests with specialized libraries~\cite{dpdk}.
Figure~\ref{fig:mssa_comparison} shows common software and hardware
architectures of CSP deployments.

The trend to replicate functionality in the userspace has two main drawbacks.
First, applications require specialized implementations of functionality that
has a standard OS interface hindering the reuse of specialized implementations.
Second, due to the added features, a larger code base offers more opportunities
to be exploited, thereby increasing the need to isolate fault domains.

\paragraph{Memory-Safe Languages}
Memory-safe languages and runtimes~\cite{wasm, nacl, rust} improve the developer
productivity and application security by eliminating memory-safety violations
such as buffer overflows~\cite{projectzero2021report}. While memory safety
guarantees vary, memory safe languages generally limit any memory access of an
application to a subset of the entire address space, and can be combined control
flow integrity to avoid bypassing or altering runtime security checks. The most
common forms are object-based memory safety which enforces an access granularity
on an object-level, whereas virtual machine-based memory safety only limits
access to the memory area defined by the virtual machine.

Recently, Rust~\cite{rust} and Webassembly (Wasm)~\cite{wasm} innovated the
state of the art in two different directions. Throughout this paper we refer to
Rust and Wasm as a representative for their class of memory-safe language. Rust
is a compiler-enforced memory-safe language and instead of relying on garbage
collection, relies on the model of borrowing and ownership of memory objects.
Recent work focuses on proving security for these unsafe
environments~\cite{rustbelt}, providing secure environments limiting the
capabilities of unsafe code~\cite{pkru-safe}, or isolating
components~\cite{isolationinrust}. The Wasm specification describes both, a byte
code and a virtual machine executing the byte code, providing security
guarantees regarding the control flow and memory boundaries. Recent work
recognizes the importance of memory-safe languages and suggests their use in a
moonshot project to build a novel software architecture with supporting
hardware~\cite{cybersecuritymoonshot}.

\paragraph{Single Address Space OS (SASOS)}

Before CPUs supported multiple address spaces and page tables, OSes were built
assuming a single address space. These SASOS relied on capability
systems~\cite{opal} and software fault isolation~\cite{singularity} to isolate
multiple applications and OS components. Most recently, Microsoft with the
Midori and Singularity OS~\cite{midori, singularity} tried to build a SASOS to
improve the execution of distributed workloads and formally verify security
guarantees of the toolchain and system. Additionally, CSPs more and more rely on
unikernels~\cite{mirageos, unikraft, graphene} to improve performance. The
efficiency of these systems lies in the near-zero cost of sharing memory and
low-cost context switching.

\paragraph{Memory-safe Library OS}

Demikernel~\cite{demikernel}, Occlum~\cite{occlum}, FaaSm~\cite{faasm}, or
cubicleOS~\cite{cubicleOS} are library OSes (library OS) reling on memory-safe
languages to reduce the attack surface of the library OS and FaaS runtimes like
FaaSm~\cite{faasm} additionally isolate functionality using Wasm. These efforts
hint towards an environment in which most software executes within a memory-safe
environment, either by being written in a memory-safe language (e.g., Rust) or
being compiled to Wasm which offers sandbox isolation guarantees.

\section{\arch}

In this section we provide an understanding of how memory-safe languages and
runtimes help building a novel software runtime environment and what hardware
optimizations and features improve performance.

\subsection{Isolating Services}

\archshort relies on memory-safe languages and runtimes to isolate memory of
services, build independent fault domains~\cite{multics}, and enable sharing to
allow efficient use of services within the system. The goal is to build these
abstractions without relying on traditional CPU capabilities such as address
space identifiers, rings, supervisor mode or privileged instructions.

To isolate memory accesses of services, \archshort relies on Software Fault
Isolation (SFI)~\cite{sfi} techniques in compilers or interpreters, or
compiler-enforced object-granular memory-safety. Since memory safety guarantees
vary between languages and runtimes, their capabilities have to be analyzed,
understood and measures taken to ensure that each service is isolated. Recent
work~\cite{isolationinrust} analyzes the memory safety properties of Rust for
isolating services.

We suggest researching a unifying abstraction layer building a foundation across
languages and runtimes providing different memory-safety properties. The
difference between Rust and Wasm, for example, lies in the ability of Rust to
enforce memory safety on a per object level as long as the type system is
respected, whereas Wasm enforces memory safety at a coarse granularity limiting
memory to the entire Wasm virtual machine memory space. Rust's type system helps
when two Rust implementations exchange data. In contrast, sharing data across
two independent Wasm modules requires new techniques to make the same memory
available to both modules at the same time. Ideally, hardware and software
mechanisms enable sharing and isolation across different languages and runtimes
allowing for efficient software-only sharing when possible and using hardware to
enforce sharing when the software mechanisms do not enforce fine-grained memory
safety.

A unified memory-safe service isolation provides the basis for private memory
in each service. In case these techniques allow unsafe execution (no memory
safety properties during the execution), the compiler needs to either prevent
their execution or further restrict access by deploying runtime bounds checks.
To avoid control flow vulnerabilities bypassing the memory-safety measures, the
compiler additionally needs to restrict and control jump and call targets to the
set of potential landing targets. Both, Rust and Wasm, achieve these goals with
no or minor changes.

\subsection{\archshort~OS or library OS}

We discuss two alternatives to realize \archshort, a memory-safe OS and a
multi-service memory-safe library OS. Our analysis considers existing work on
building Rust-based OSes~\cite{buildingrustos, caseforrustos, tockrustkernel,
redox}. Building a new OS provides the greatest possible flexibility and avoids
some of the existing inherent overheads that traditional OSes bring. Its
downside is the additional implementation overhead and a long path towards
adoption. Alternatively, a library OS implementation allows adoption in an
existing environment such as Linux. Due to the dependence on a host OS, not all
of the hardware, abstraction and scheduling optimizations can be implemented
with a library OS limiting the possible performance gain.

The main difference between the two alternatives, is how the \archshort OS
handles interrupts, timers, and other low-level traditional OS functionality by
itself, whereas the library OS needs to communicate with the hosting OS to
achieve these behaviors. A prominent example is the control over page tables
which is impossible without changes to a traditional OS from the userspace. Page
table-based optimizations will be much harder with a library OS-based approach and
their efficient setup or modification could limit performance benefits.

\paragraph{System Calls and interactions}

Traditionally, an application accesses and shares information with the host or
network-connected services via the system call interface. Our goal is to provide
an extensive interface resembling dynamic library loading rather than fixed
system calls. To allow memory-safe services to exchange information, the memory
access capabilities have to be transferred. When two Rust services communicate,
they could agree on the same underlying type system of the exchanged memory and
continue to adhere to the memory safety properties. Similarly, Rust's ownership
model lends itself to this architecture to borrow the output of one service to
another. To keep track of this borrowing, a runtime service has to keep track of
the most recent owner to ensure proper destruction of the memory once it is no
longer in use.

Today memory-safe languages do not support sharing types systems across
service domains. Each language would have to be extended to support this
type system. With languages like Wasm, such cross-service communication is
not protected by the memory safety properties and even breaks the memory access
model of Wasm virtual machines which assume a bounded memory size.

To overcome these limitations, we suggest a software-only based technique which
offers proxy access. A proxy access allows a service to switch to a function
with access to the shared memory object (e.g., via a list of proxy functions
accessible to the function) while not switching the service's context. As a
result, access to the shared memory is as quick as a function call, and such
capabilities can be arbitrarily granted by generating the according proxy access
function. A proxy-aware compiler could generate proxy functions and
translate memory accesses into proxy functions calls to transparently switch
between the two.

\paragraph{Common Services}

Both alternatives have to provide common OS functionalities such as a file
system, network connections, and per service specific metadata such as file
descriptors. To allow specialization for workloads, only a minimal
substrate should be provided, and dependencies should be selected by each
service. The initially needed building blocks are a service loader to enable
execution of services, and an interface for each service to discover
other services providing certain functionality. In this regard, discovering
a service is possible by name and leads to loading a function pointer table into
the execution of the service. Any subsequent communication happens via this
function table instead of calling into the OS via system call. As a result, a
system call becomes a function call instead. All the service switching
mechanics, if needed, are hidden within the function call pointers as prepared
by the only real system call to discover other services. This design offers
the most flexibility, like microkernels~\cite{microkernel, microkernel2},
and offers the ability to build an abstraction layer for legacy applications
mapping the new interface to previous interfaces when needed.

As part of the discovery service, the system negotiates between different memory
safety properties, take advantage of fine-grained properties, or rely on
hardware-provided capabilities to allow efficient access. For instance, assuming
all services and services were built in Rust and the supply chain would
provide evidence of this fact, the system could rely on Rust's internal type
system to protect access capabilities when services exchange information.

\subsection{Hardware implications}

As discussed in previous sections, some of the traditional hardware
functionalities can alternatively be provided by relying on memory safety
instead. Consequently, \archshort offers several avenues to further optimize
hardware for improved performance and security.

\paragraph{Address translation and caches}

Traditional address translation in CPUs relies on page tables and caches within
the CPU to reduce the number of page walks. The translation lookaside buffer
(TLB) caches recent translations between virtual and physical memory. In case a
miss occurs, the page miss handler (PMH) walks the actual page table to find a
possible mapping. This takes several hundred cycles to complete and requires
memory bandwidth to receive the page table entries from memory. Services
with high memory bandwidth demands particularly suffer when frequent page walks
occur which reduce the memory bandwidth~\cite{rebootingmemory}.

Wasm simplifies the memory model in its virtual machine specification to a
linearly growing memory area. For cloud services, the required memory size is
typically known before launch and described in deployment files. This simplified
memory layout with its allocations and mappings can be coalesced into larger
regions and ideally mapped such that their address translation can be performed
statically. Such a static mapping removes entries from the TLB lowering the
pressure for most memory accesses and avoids any page table walks, since the
mapping is statically configured. To avoid fragmentation, different power of 2
buckets could be created in physical space to allocate different maximum memory
sizes. Additional research is needed to adopt these optimizations to
memory-safe-only environments and further improve the efficiency and performance
of hardware mechanisms.

\paragraph{Hardware-support for efficient memory sharing}

In \archshort memory-safe languages restrict memory accesses of a service and do
not allow arbitrary sharing between services limiting the efficiency of
communication. To overcome this restriction, hardware and software techniques
can provide efficient ways to communicate. Traditional OSes solve this problem
via shared memory, but \archshort shares all virtual memory automatically. As a
result, only the memory-safety guarantees restrict accesses to arbitrary memory.
We can ideally design techniques to safely open this restrictive design while
maintaining the performance and security guarantees that memory-safe languages
provide.

Hardware can assist the memory safe language by providing low level access
capabilities to allow sharing of memory when the memory safe language does not
allow such fine-grained access permissions like Wasm or in case of Rust when the
type system is not known at compilation time.

Existing work on process-based isolation~\cite{codoms, dipc, coffer}, provides a
capability-based system based on page tables. In these systems, fine grained
memory sharing is enabled via specially crafted memory capabilities allowing
multiple processes to share access to memory with other processes.

To achieve similar sharing within the same virtual address space,
CHERI~\cite{cheri} or Cryptographic Computing~\cite{cryptocompute} provide the
ability to set capabilities from within the same virtual address space without
involvement of the OS.

\paragraph{Constraining legacy applications}

\archshort adoption is limited by implementing new or reimplementing existing
services for this architecture. Translating existing services automatically
allow faster adoption, but requires them to become memory safe. The service
needs to be restricted to the service' memory and control flow boundaries.

Wasm can be used as compilation target in common compilers today and transform
existing applications to a virtual machine definition limiting access of the
application into a 4GB memory and providing a well-structured control flow
sufficiently restricting the application to be used in \archshort system.
Unfortunately, Wasm degrades performance compared to native and requires access
to source code.

In contrast to such a software-only approach, are hardware-based
techniques~\cite{erim, hodor, donky, secage, endokernel, galeed} (mostly \intel
MPK-based) which rely on CPU features to restrict memory accesses.
Unfortunately, these systems require elaborate security
monitors~\cite{endokernel} and the hardware features may not be readily
available. As a result, novel architecture extensions could help translate
legacy applications into \archshort.

\section{Use Case: Microservices and Service Meshes}

\archshort improves the efficiency of cloud workloads. These workloads benefit
from the efficient resource sharing, specialization at the OS-level or
efficiently bypassing the OS.

Several cloud services are implemented as microservices, a set of functionally
independent, but highly connected components. These components are deployed via
an orchestration framework and connected via a service mesh governing access
between different components and allowing to observe the behavior of the
service. Existing frameworks deploy such workloads via containers and virtual
machines completely isolating each component from each other. In addition,
service meshes~\cite{istio} deploy proxies like envoy~\cite{envoy}, intercepting
all network communication in a co-located container. As a result network
messages have to be copied between containers and OS multiple times and objects
marshaled.

To avoid this overhead, \archshort deploys service mesh proxies and components
in the same address space allowing them to communicate and invoke each other via
shared interfaces. In addition, if components are co-located their communication
can be short-circuit as well, and a service mesh proxy could even eliminate
itself from the communication path. As a result, we copy network messages less
and marshalling/unmarshalling may not be necessary. Netbricks~\cite{netbricks}
suggests similar optimizations for network functions.

\section{Discussion}

\archshort relies on memory-safe languages instead of traditional hardware
functionality and as a result, offers several avenues to optimize hardware for
improved performance and security. This direction reduces the reliance on
long-used and battle-proven isolation techniques. Memory-safe languages and
runtimes provide certain security properties, but their capabilities are limited
with respect to hardware-based attacks and protecting against them may incur
about 2x runtime overhead~\cite{narayan2021swivel}. CHERI~\cite{cheri} and
Cryptographic Computing~\cite{cryptocompute} offer capability-based replacements
that \archshort could rely on to strengthen the security properties.
Alternative, and more light-weight approaches should be considered like bringing
back 32-bit segmentation or the RISC-V J extension~\cite{riscvj}. Recently,
hardware-based light-weight subprocess isolation has been an active area of
research with promising techniques like Donky~\cite{donky} suggesting a RISC-V
extension similar to \intel MPK or research using \intel MPK for
isolation~\cite{erim, hodor, endokernel}. Additional exploration and research in
this hardware and software co-design can bridge the gap and overcome today's
performance and security limitations.

The security foundation of \archshort depends on the memory-safe compilation
tool chain and a small, but important runtime component. The runtime component
needs to load services within the single address space while respecting the
security requirements of the memory-safe language's toolchain. Each toolchain
may have different requirements regarding the memory layout (e.g., Wasm requires
an 8 GB space around each module's heap). Some toolchains, e.g. Webassembly,
lends itself to runtime verification allowing to check binary code for memory
safety guarantees before starting to execute it~\cite{veriwasm}. Another path to
establish trust in each toolchain is to vet and certify it, and securely record
the supply chain of the service. Before starting a service the runtime would
authenticate the trusted toolchain and the metadata of a
service~\cite{marcelasecureorch}. Alternatively,
proof-carrying-code~\cite{necula1997proof} can establish trust in the binary of
a service. To further strengthen the security and reduce the dependencies,
future research should explore formal verification and trusted supply chains.

Several research prototypes~\cite{faasm,lwc, erim, narayan2021swivel} run
services in the same address space to improve communication overhead.
Fastlane~\cite{fastlane} is the first to automatically combine multiple services
running inside containers to be combined into a single container. Their
prototype efficiently combines Python-based services and carefully parallelizes
service invocations to improve performance. While Fastlane proposes an
interesting direction to improve the current environment, it is limited to a
single language, weaker hardware isolation with limited security guarantees. In
contrast, \archshort focuses on the use of memory-safe languages and leveraging
hardware optimizations for optimal performance and security tradeoff between
hardware and software.
\section{Conclusion}
Cloud workloads have shifted towards deploying small functional units heavily
communicating within a single machine or across a set of machines leading to an
unsustainable infrastructure tax of about 25\%. In this paper, we hypothesize a
\arch to reduce these overheads and generalize the environment to a larger set
of applications. We argue that existing languages and runtimes provide
sufficient security guarantees to implement a \archshort and take advantage of
the benefits of memory-safe languages and runtimes for these workloads. We
highlight open research questions and potential optimizations to further improve
the architecture.

\begin{acks}
We would like to thank the reviewers of SecDev'22, WORDS'22,
Nathan Dautenhan, Fangfei Yang, Shravan Narayan, Mansour Alharthi and several
Intel colleagues for their insightful feedback and discussion.
\end{acks}

\bibliographystyle{plain}
\bibliography{biblio}

\end{document}